\documentclass[a4paper, reqno, 11pt]{amsart}

\usepackage{amsmath,amsthm,amssymb,color,graphicx}
\usepackage{amscd,verbatim}
\usepackage[all]{xy}

\theoremstyle{plain}
\newtheorem{theorem}{Theorem}
\newtheorem{proposition}[theorem]{Proposition}
\newtheorem{lemma}[theorem]{Lemma}
\newtheorem{corollary}[theorem]{Corollary}

\theoremstyle{definition}
\newtheorem{definition}[theorem]{Definition}

\theoremstyle{remark}
\newtheorem*{remark}{Remark}

\numberwithin{equation}{section}
\numberwithin{theorem}{section}
\numberwithin{figure}{section}
\numberwithin{table}{section}
\numberwithin{example}{section}

\newcommand{\beq}{\begin{eqnarray}}
\newcommand{\eeq}{\end{eqnarray}}

\newcommand{\cD}{{\mathcal D}}

\newcommand{\cH}{{\mathcal H}}

\newcommand{\cK}{{\mathcal K}}
\newcommand{\cL}{{\mathcal L}}

\newcommand{\bbC}{{\mathbb C}}

\newcommand{\RR}{{\mathbb R}}
\newcommand{\bbR}{{\mathbb R}}
\newcommand{\R}{{\mathbb R}}

\newcommand{\bbT}{{\mathbb T}}
\newcommand{\ZZ}{{\mathbb Z}}
\newcommand{\bbZ}{{\mathbb Z}}
\newcommand{\Z}{{\mathbb Z}}

\DeclareMathOperator{\Hom}{Hom}

\def\ip#1#2{\langle{#1},{#2}\rangle}
\def\dip#1#2{\langle\kern-2pt\ip{#1}{#2}\kern-2pt\rangle}

\def\sect#1{\advance\count30 by 1\count31=0 
\bigskip\noindent{\bf \the\count30. #1}

\bigskip} 
\def\dfn{\advance\count32 by 1
\bigskip\noindent{\bf Definition \the\count30.\the\count32.  }}
\def\thm#1{\advance\count31 by 1
\bigskip\noindent{\bf #1 \the\count30.\the\count31.  }}
\def\athm#1{\advance\count31 by 1
\bigskip\noindent{\bf #1 A.\the\count31.  }}

\begin{document}

\title[Harmonic Cheeger-Simons characters]
{Harmonic Cheeger-Simons characters with applications}

\author[R. Green]{Richard Green}

\address[Richard Green]{
Department of Pure Mathematics,
University of Adelaide, 
Adelaide 5005, 
Australia}
\email{ric.green@adelaide.edu.au}

\author[V. Mathai]{Varghese Mathai}

\address[Varghese Mathai]{
Department of Pure Mathematics,
University of Adelaide, 
Adelaide 5005, 
Australia}
\email{mathai.varghese@adelaide.edu.au}

\begin{abstract}
We initiate the study of harmonic Cheeger-Simons characters,
with applications to smooth versions of the Geometric Langlands program in the 
abelian case. 
\end{abstract}


\maketitle


\section{Introduction}
Cheeger-Simons characters, also known as differential characters,  provide a useful refinement of integral cohomology incorporating differential forms. The original construction and basic properties are summarized in \cite{CS}, which also details corresponding refined characteristic classes and associated applications. Such refinements of cohomology theories are now known as differential cohomology theories. A recent, elegant treatise on the subject is \cite{HS}, which  includes a construction of differential versions of arbitrary generalized cohomology theories, together with interesting applications, such as to quadratic refinements of the intersection pairing on a smooth manifold.
An application to abelian gauge theories, the simplest example being electromagnetism, can be found in \cite{FMS},  where it is shown that in the quantum theory, fluxes may not be simultaneously measurable.
There have also been numerous other applications, either of Cheeger-Simons characters
or of its equivalent, Deligne cohomology (cf. \cite{Br}), both in mathematics and mathematical physics.
 
For  a compact Riemannian manifold $X$, we introduce the notion of 
{\em harmonic Cheeger-Simons characters} in Definition \ref{defn:harmonicCScharacters},
and establish the fundamental exact sequences that it satisfies in Lemma \ref{lem:exact}
as well as some of the basic properties such as Poincar\'e-Pontrjagin duality in subsection \ref {subsect:PD}. 
The group of all harmonic Cheeger-Simons $j$-characters $\check{\cH}^j(X)$ form a subgroup of 
the group of all Cheeger-Simons $j$-characters, and a crucial point established 
here is that  $\check{\cH}^\bullet(X)$ is {\em finite dimensional}, and can be viewed as a 
refinement of the space of harmonic
forms on $X$, as it also encodes the torsion subgroup of $H^\bullet(X, \bbZ)$.
In fact, we establish an analogue of the Hodge theorem in this context in
Proposition \ref{prop:harmonicCShodge}.
We also compute these groups for several examples in this 
section, such as for compact connected Lie groups endowed with a bi-invariant 
Riemannian metric
in Lemma \ref{lemma:harmchi}.

In the last section, we introduce the {\em harmonic Picard variety} for compact Riemann surfaces, 
which is fundamental for 
several constructions related to the smooth analogue of the Geometric Langlands program in the 
abelian case, as established in Theorem \ref{thm:geomHecke}. The Geometric Langlands program 
is currently of central importance in gauge theory and string theory 
due to the relationship with S-duality in supersymmetric gauge theory 
in four dimensions. Although still a conjecture in the general
nonabelian case, recently several important insights have been given by E. Witten and his
collaborators, in a series of long papers starting with \cite{KW}. \\

\noindent{\bf Acknowledgements.} R.G. acknowledges the receipt of an 
Australian Postgraduate Award. V.M. acknowledges 
support from the Australian Research Council.

\section{Preliminaries}

Here we recall for convenience some classical results on line bundles over 
surfaces and connections on these, whose proofs are included for the convenience of the reader. Also 
included are sections recalling results on symmetric products of surfaces and characters of the 
fundamental group, which is mainly used in \S\ref{sect:picard}.

\subsection{Divisor line bundles on oriented surfaces}
Let $X$ be an oriented surface, and $p\in X$.
Then we can cover $X$ with two open sets,
$U_1\cong \R^2$ a coordinate neighbourhood of $p$, and
 $U_0=X\backslash \{p\}$. Then
 $U_0\cap U_1\cong \R^2\backslash \{0\}\cong S^1\times \R.$
 We define a complex line bundle $\cD_p$ over $X$, called the {\em divisor line bundle}
 associated to $p\in X$, by taking the transition function $g_{01}$ on
$U_0\cap U_1$ to be the pull-back from $S^1$ of an $S^1$-valued 
smooth map of degree one, whose homotopy class is the generator of $H^1(S^1,\Z)$.
The choice of orientation on $X$ induces an orientation on $S^1$ and hence 
a choice of generator. Since there are only two open sets in this definition, 
nothing further needs to be checked.
If $X$ is compact, then the first Chern class, $c_1(\cD_p) \in H^2(X,\Z)\cong \Z$
is the generator, and a refinement of this 
statement will be discussed in the sequel. This construction is a 
smooth analogue of the well known construction of a holomorphic line bundle 
 associated to a point  in a Riemann surface.

\subsection{Connections with harmonic curvature on divisor line bundles}
Now suppose that $X$ is a {\em compact} oriented surface. Choose a Riemannian
metric $g$ on $X$ with volume form $V$ and  total volume $2\pi$. Then $V$ is 
a harmonic 2 form, 
which represents the image of the generator of $H^2(X, \Z)$ in $H^2(X,\R)$.
We will now construct a connection, $\nabla_p$, with curvature $V$, on the 
divisor line bundle $\cD_p$ associated to a point $p \in X$.

We shall use the language of currents, or distributional forms. On an
n-manifold $M$, a smooth $k$-form $\alpha$ defines a linear form on
$\Omega^{n-k}(M)$ by
$$\langle \alpha, \beta\rangle=\int_M\alpha \wedge \beta$$
and then by Stokes' theorem
$$\langle d\alpha, \beta\rangle=(-1)^{k+1}\langle \alpha, d\beta\rangle.$$
The theory of de Rham currents considers general continuous linear functionals on
$\Omega^{n-k}(M)$, rather than just those coming from k-forms as above.
In particular,
an example in the case $k=n$ is the Dirac delta function
$\delta_p$ associated to a point $p\in M$, defined as $\delta_p(\phi) = \phi(p)$. 
There is also an analogous Hodge theory of currents (cf \cite{dR}).

The homology class of a point $p\in X$ is dual to the de Rham cohomology class of
$V/2\pi$. In terms of currents, the
degree 2 current $V-2\pi \delta_p$ is null cohomologous.
It follows from the Hodge theory of currents that there is a 
degree 2 current $H_0$
such that
$$\Delta H_0=V-2\pi\delta_p.$$
The current $H_0$ is unique up to the addition of a constant
multiple of $V$, and by elliptic regularity  $H_0$ is a 2-form which is
smooth except at $p$ where the function $\phi=\ast H_0$ has a singularity of
the form $\phi= {\rm constant}.\log(r)+\dots$
Choose a local smooth 2-form $H_1$ on the coordinate neighborhood $U_1$
such that $\Delta H_1=V$
and define the 1-forms $F_0=d^*H_0$ on $U_0=X\backslash\{p\}$ and
$F_1=d^*H_1$ on $U_1$. Note that $F_0$ is now independent of the choice of
$H_0$. Then $dF_0=dd^*H_0=\Delta H_0=V$ on $U_0$ and $dF_1=dd^*H_1=V$ on $U_1$,
therefore $d(F_1-F_0) =  0$ on $U_0\cap U_1$. 
Hence, the locally defined 1-forms  $\{F_1, F_0\}$ will patch together to give a connection 
on the line bundle $\cD_p$ with curvature $V$, provided $(F_1-F_0)/2\pi= d \log (g_{0 1})$ on 
the overlap $U_0\cap U_1\cong \R^2\backslash \{0\}$ determines a cohomology
class which is the generator of $H^1(\R^2\backslash \{0\})\cong \Z$. Consider
a closed ball $B$ centred at $p$ in the coordinate neighborhood $U_1$
and $\varphi$ a smooth function of compact support in $U_1$ which is
identically 1 in a neighborhood of $B$. By the definition of $H_0$ and $H_1$,
$$\langle dd^*H_0, \varphi\rangle=\int_{U_1}\varphi V-2\pi\varphi(p)
\qquad \text{and} \qquad 
\int_{U_1} \varphi dd^*H_1=\int_{U_1}\varphi V,$$
so subtracting,
$$\langle d(d^*H_0-d^*H_1),\varphi\rangle =-2\pi$$
since $\varphi\equiv1$ near $p$. Since $d^*H_0-d^*H_1$ is a smooth  closed 1-form
outside $p$, 
so we can equivalently consider its restriction to $B$, and since
$\varphi=1$ on $B$,  we see by Stokes theorem that,
$$-2\pi=\langle  d(d^*H_0-d^*H_1), \varphi \rangle=
\int_B d(d^*H_0-d^*H_1)= \int_{\partial B}(d^*H_0-d^*H_1).$$
Thus  $(F_1-F_0)/2\pi=d^*(H_1-H)/2\pi$ determines an integral cohomology class, 
which is a generator of the cohomology group as asserted. Thus we have produced a
connection on the divisor line bundle $\cD_p$, represented by the pair of local 1-forms
$\{F_1, F_0\} = \nabla_p$, whose curvature is equal to the harmonic 2-form $V$.

\subsection{Symmetric products of spaces}

Let $M$ be smooth manifold. On the n-fold Cartesian product $M^n$, one can define an action of the 
symmetric group $S_n$  as follows. Recall that $S_n$ consists of bijections 
$\theta : \{1, 2, \dots , n\} \to \{1, 2, \ldots , n\}$ (also called permutations of $n$ letters)
and has order equal to $n!$, where $S_2= \bbZ_2$ and $S_n$ is nonabelian when $n>2$. 
The action of $S_n$ on $M^n$ is given by  
$$ 
\begin{array}{rcl}
M^n \times S_n & \longrightarrow & M^n \\[7pt]
((x_1, \ldots, x_n), \theta)  & \longrightarrow & (x_{\theta(1)}, \ldots,  x_{\theta(n)}) 
\end{array}
$$
The quotient $M^n/S_n$ is an {\em orbifold} denoted by ${\sf Sym}^{(n)}(M)$ 
and is called the $n$-th symmetric product of $M$.  
It consists of all the unordered $n$-tuples of points in $M$.

 In the special case when  $\dim(M)=2$, the symmetric product 
${\sf Sym}^{(n)}(M)$ is not just an orbifold, but is a manifold; details
 can be found  in \cite{GH}, page 236. The key point is the observation
 that ${\sf Sym}^{(n)}(\bbC)$ is diffeomorphic to $\bbC^n$ via the diffeomorphism $\phi$
 given by $\phi(z_1, \ldots z_n) = (\sigma_1, \ldots \sigma_n)$, where $\sigma_j, \, j=1, \ldots , n$
 are the elementary symmetric functions in $z_1, \ldots , z_n$, that is, the coefficients of the 
 polynomial $p(z)= \prod_{j=1}^n (z-z_j)$.
 
 For example, ${\sf Sym}^{(n)}(\bbC P^1) \cong \bbC P^n$, which can be deduced using
the Schubert cell decomposition of $\bbC P^1$ and $\bbC P^n$. Also,
 ${\sf Sym}^{(n)}(\bbT^2)$ is a fiber bundle over $\bbT^2$ with 
fiber diffeomorphic to $\bbC P^{n-1}$, cf. \cite{Mattuck}. This description enables 
one to determine many useful facts about the topology of  ${\sf Sym}^{(n)}(\bbT^2)$.

The product map $M^n \times M^m \mapsto M^{n+m}$ descends to a continuous map 
of the quotient spaces 
\begin{equation}\label{eqn:symm}
h_{m,n} : {\sf Sym}^{(m)}(M) \times {\sf Sym}^{(n)}(M)
\to {\sf Sym}^{(m+n)}(M),
\end{equation} 
and is given by $([x_1, \ldots, x_m], [y_1, \ldots ,y_n])
\mapsto [x_1, \ldots, x_m, y_1, \ldots , y_n]$.

Choosing a point $p \in M$ determines an embedding
$M^n \hookrightarrow M^{n+1}$ given by $(x_1, \ldots, x_n) \to (x_1, \ldots, x_n, p)$,
and the embedding ${\sf Sym}^{(n)}(M) \hookrightarrow {\sf Sym}^{(n+1)}(M)$
given by $[x_1, \ldots, x_n] \to [x_1, \ldots, x_n, p]$. Define 
$$
 {\sf Sym}^{(\infty)}(M) = \bigcup_{n\ge 1}  {\sf Sym}^{(n)}(M)
$$
in the weak topology. It is a commutative, associative H-space with a strict identity
element. The Dold-Thom theorem (cf \cite{AH}) asserts that there is a homotopy equivalence
\begin{equation}\label{eqn:Dold-Thom}
 {\sf Sym}^{(\infty)}(M) \sim \prod_{n\ge 1} K(H_n(M, \bbZ), n)  
\end{equation}
where the 
right hand side denotes the cartesian product of Eilenberg-Maclance spaces.
In particular, when $M$ is a compact Riemann surface of genus equal to $g$, one has the homotopy equivalence
$$
 {\sf Sym}^{(\infty)}(M) \sim  {\mathbb T}^{2g}  \times \bbC P^\infty
$$

If $f: M \to N$ is a continuous map, then it induces in a natural way a map on the cartesian products
$f^n : M^n \to N^n$, which descends to a continuous map 
$ {\sf Sym}^{(n)}(f) :  {\sf Sym}^{(n)}(M) \to  {\sf Sym}^{(n)}(N)$. If $f, g : M \to N$ 
are continuous maps that are homotopic, then so are the continuous maps
$ {\sf Sym}^{(n)}(f)$ and $ {\sf Sym}^{(n)}(g)$. In particular, if there is a homotopy
equivalence $M \sim N$, then there is a homotopy equivalence
$  {\sf Sym}^{(n)}(M) \sim  {\sf Sym}^{(n)}(N)$.

\subsection{Characters of the fundamental group and of the first homology group}
Suppose $M$ is path connected. Then the  Hurewicz theorem gives a canonical homorphism 
$\phi: \pi_1(M) \to  H_1(M, \bbZ)$, which is surjective, and has kernel the commutator subgroup $[\pi_1(M), \pi_1(M)]$
of $\pi_1(M)$. As such, this homomorphism descends to an isomorphism between the abelianization
of  the fundamental group, $\pi_1(M)_{ab} \cong H_1(M,\ZZ)$.
Since every character of $\pi_1(M)$  factors through its abelianization, it induces a character 
of $H_1(M, \bbZ)$. Conversely, every character of $H_1(M, \bbZ)$ pulls back via $\phi$ to a 
character of $\pi_1(M)$. This sets up a bijective correspondence between characters of 
$\pi_1(M)$ and characters of $H_1(M, \bbZ)$. The following proposition, which will be used
in \S4, is due to \cite{AG} and asserts in particular the surprising fact 
 that the fundamental group of  symmetric 
products of a Riemann surface, is abelian.

\begin{proposition}\label{prop:fund-gp}
If $M$ is a Riemann surface and $n > 1$, then 
$$\pi_1 ({\sf Sym}^{(n)}(M)) \cong H_1 (M, \bbZ ),$$
\end{proposition}

\begin{corollary}
The characters of $\pi_1(M)$ are in bijective correspondence
with the characters of $\pi_1({\sf Sym}^{(n)}(M))$. 
\end{corollary}

\section{Harmonic Cheeger-Simons characters}

We begin by reviewing the definition  and some basic properties
of Cheeger-Simons characters on a smooth manifold $X$, cf. \cite{CS, FMS, HS}. 
If $X$ is a compact Riemannian manifold, then we introduce the notion of 
{harmonic Cheeger-Simons characters} in Definition \ref{defn:harmonicCScharacters},
and establish the basic properties in Lemma \ref{lem:exact} 
and Proposition \ref{prop:harmonicCShodge}. 

\begin{definition}
A {\em Cheeger-Simons $j$-character} is an element
$\chi \in \Hom(Z_{j-1}(X), \bbR/\bbZ)$ such that there is a 
$j$-form $F_\chi$ on $X$ with the property that if $\Sigma
\in B_{j-1}(X)$, i.e $\Sigma = \partial Q$, then
$$
\chi(\Sigma) = \int_Q F_\chi \quad \mod \bbZ.
$$
\end{definition}

This automatically implies that $F\chi$ is a closed $j$-form, with integral periods, which we shall write as $F_\chi \in \Omega^j_{\ZZ}(X)$.
 The form $F_\chi$ is uniquely determined by the character $\chi$, and referred to as its fieldstrength
 or curvature.
The group of Cheeger-Simons $j$-characters is denoted by $\check{H}^j(X)$,
where the group law is given by pointwise sum of Cheeger-Simons characters.
We often also denote elements 
in $\check{H}^j(X)$ by $[\check{A}]$, where $\check{A}$ is a locally defined $(j-1)$-form on $X$ 
sometimes called a vector potential, such that 
$d\check{A}$ is the globally defined form given by the fieldstrength.

In addition to the mapping $F: \check{H}^j(X) \rightarrow  \Omega_{\ZZ}^j(X)$ given by the fieldstrength, there is a mapping $c: \check{H}^j(X) \rightarrow H^j(X,\ZZ)$, called the characteristic class map, for which the  following diagram commutes 
\begin{eqnarray*}
\xymatrix{
 \check{H}^j(X) \ar[d]_c \ar[r]^{F} & \Omega_{\ZZ}^j(X)\ar[d]  \\
  H^j(X,\ZZ) \ar[r]& H^j(X,\RR).
     }
\end{eqnarray*}
Here the rightmost arrow is given by taking de Rham cohomology classes, and the bottom arrow is
induced by the inclusion $\ZZ \rightarrow \RR$.

There are two fundamental exact sequences associated to $\check{H}^j(X)$. The first is related to the fieldstrength and given by
\begin{equation}\label{eqn:ses1}
0 \to H^{j-1}(X, \bbR/\bbZ) \to \check{H}^j(X) \stackrel{F}{\to} \Omega^j_\bbZ(X) \to 0.
\end{equation}
The other is related to the characteristic class and given by
\begin{equation}\label{eqn:ses2}
0 \to \Omega^{j-1}(X)/\Omega^{j-1}_\bbZ(X) \to \check{H}^j(X) \stackrel{c}{\to}  H^j(X, \bbZ) \to 0.
\end{equation}
These exact sequences give a picture of the structure of the Cheeger-Simons 
cohomology groups, viz that they are infinite dimensional, with components labelled by 
$H^j(X, \bbZ) $ and each component is a torus bundle with typical
fiber $ H^{j-1}(X, \bbZ) \otimes \bbR/\bbZ$, over the infinite dimensional vector space 
$\Omega^{j-1}(X)/\Omega^{j-1}_{cl}(X)$, where $\Omega^{j-1}_{cl}(X)$ denotes closed forms.

In particular, the exact sequences \eqref{eqn:ses1} and \eqref{eqn:ses2} show that the Cheeger-Simons
groups $\check{H}^\bullet(X)$ can be viewed as a refinement of the de Rham cohomology of $X$, 
as it also encodes the torsion subgroup of $H^\bullet(X, \bbZ)$. 

If $X = (X,g)$ is a compact Riemannian manifold, then the space of harmonic $j$-forms
$\mathcal{H}^j(X)$ can be defined and is, by the Hodge theorem, isomorphic to $H^j(M,\RR)$, while 
the harmonic $j$-forms with integral periods $\mathcal{H}_{\ZZ}^j(X)$ form a lattice isomorphic to the image of $H^j(M,\ZZ)$ in $H^j(M,\RR)$. In this situation we will define as follows, a finite dimensional subgroup of the 
full Cheeger-Simons character group which encodes the same topological information.

For the remainder of this section, let $X$ be a compact Riemannian manifold.The following is 
the main new definition that we introduce in the paper. 

\begin{definition}\label{defn:harmonicCScharacters} 
A {\em harmonic Cheeger-Simons $j$-character} is a character 
$\chi \in \check{H}^j(X)$ whose fieldstrength $F_\chi$ is a harmonic differential form.
\end{definition}

\begin{remark}
The harmonic Cheeger-Simons $j$-characters form a subgroup of $\check{H}^j(X)$ which we denote by 
$\check{\cH}^j(X)$. The crucial point established in the following Lemma is that  
$\check{\cH}^\bullet(X)$ is finite dimensional, and can be viewed as a refinement of the space of harmonic
forms on $X$, as it also encodes the torsion subgroup of $H^\bullet(X, \bbZ)$. Moreover, it has many 
properties analogous to the full Cheeger-Simons character group. 
\end{remark}

\begin{lemma}\label{lem:exact} 
There are two fundamental short exact sequences associated to $\check{\cH}^j(X)$, namely a fieldstrength sequence,
\begin{equation}\label{eqn:ses3}
0 \to H^{j-1}(X, \bbR/\bbZ) \to \check{\cH}^j(X) \stackrel{F}{\to} \cH^j_\bbZ(X) \to 0,
\end{equation}
and a characteristic class sequence,
\begin{equation}\label{eqn:ses4}
0 \to \cH^{j-1}(X)/\cH^{j-1}_\bbZ(X) \to \check{\cH}^j(X) \stackrel{c}{\to} H^j(X, \bbZ) \to 0.
\end{equation}
\end{lemma}

\begin{proof} 
The restriction of $F : \check{H}^j(X) \rightarrow \Omega_{\ZZ}^j(X)$ to $\check{\mathcal{H}}^j(X)$, $F_{res}  : \check{\mathcal{H}}^j(X) \rightarrow \mathcal{H}^j_{\ZZ}(X)$, is by definition surjective. Since $\ker(F)  \subseteq \check{\mathcal{H}}^j(X) $ it follows that
$\ker( F_{res} ) = \ker (F) \simeq  H^{j-1}(X,\RR/\ZZ)$.
From which we deduce the exact sequence \eqref{eqn:ses3} immediately.

The restricted map $c_{res} : \check{\mathcal{H}}^j(X) \rightarrow H^j(X,\ZZ)$ can be seen to be onto as follows. Consider a class $[u] \in H^j(X,\ZZ)$. Then by the Hodge theorem, there is a harmonic form $h \in \mathcal{H}^j(X)$ such that $[u] = [h]$ as real cohomology classes, and hence $h - u = \delta T$ for some cochain $T \in C^{k-1}(M,\RR)$. Restricting $T$ to cycles $Z_{k-1}(X)$ and reducing modulo $\ZZ$ gives the required  harmonic character $\chi = \tilde{T} \vert_{Z_{k-1}}$ with characteristic class $c(\chi) = [u]$.

The kernel of $c_{res}$ can be computed by considering the map \[\varphi: \Omega^{j-1}(X) / \Omega_{\ZZ}^{j-1}(X) \rightarrow \check{H}^j(X)\] appearing in \eqref{eqn:ses2}, which maps onto $\ker (c)$. This map $\varphi$ takes $\alpha \in \Omega^{j-1}(X)$ (viewed as a cochain via integration over chains) to the differential character $f_\alpha = \tilde{\alpha} \vert_{Z_{j-1}}$ which has curvature $d\alpha$. The Hodge decomposition implies that closed forms split as $\Omega_{cl}^j(X) = B^{j}(X)\oplus \cH^j(X)$, where $B^j(X)$ denotes exact forms. It then follows that $\varphi^{-1}(\ker(c) \cap \check{\mathcal{H}}^j(X)) = \Omega_{cl}^{j-1}(X)/\Omega_{\ZZ}^{j-1}(X)$.  Finally, the inclusion $\mathcal{H}^j(X) \rightarrow \Omega_{cl}^j(X)$ induces, via the Hodge theorem, an isomorphism $\mathcal{H}^j(X)/\mathcal{H}_{\ZZ}^j(X) \rightarrow \Omega_{cl}^j(X)/\Omega_{\ZZ}^j(X)$, so that  the exactness of \eqref{eqn:ses4} follows.

\end{proof}

The analogue of the {\em Hodge theorem} in this context is the following.

\begin{proposition}\label{prop:harmonicCShodge}
The inclusion $i: \check{\cH}^j(X) \hookrightarrow \check{H}^j(X)$ is a homotopy equivalence
(and is in fact a deformation retraction).
\end{proposition}

\begin{proof}
From the short exact sequences \eqref{eqn:ses1} and  \eqref{eqn:ses3}, we see that the quotient group
$\check{H}^j(X)/\check{\cH}^j(X) $ is isomorphic to the quotient 
$\Omega^j_\bbZ(X)/\cH^j_\bbZ(X)$. 
Since $B^j(X)$ is a subgroup of $\Omega_{\ZZ}^j(X)$, the Hodge decomposition implies $\Omega^j_{\ZZ}(X) = B^j(X)\oplus \cH^j_{\ZZ}(X)$.
Hence the quotient $\Omega^j_\bbZ(X)/\cH^j_\bbZ(X)$ is 
isomorphic to $B^j(X)$, which is contractible, and the assertion 
follows.
\end{proof}

We now give some examples of the group of harmonic Cheeger-Simons characters.

\begin{enumerate}
\item When $X=pt$, the only nontrivial groups of harmonic Cheeger-Simons characters are $\check{\cH}^0(pt)
= \bbZ$ and $\check{\cH}^1(pt) = \bbR/\bbZ$.
\item If $X$ is a compact, connected oriented Riemannian manifold of dimension equal to $N$, then 
$\check{\cH}^0(X)= \bbZ$ and $\check{\cH}^{N+1}(X) \cong \bbR/\bbZ$. More precisely, the isomorphism is 
defined as follows. Any class in 
 $\check{\cH}^{N+1}(X)$ is topologically trivial, hence any vector potential is a globally defined
 form $\check{A} \in \Omega^N(X)$, and the integration homomorphism 
  $\int_X^{\check{H}} :  \check{\cH}^{N+1}(X) \to \bbR/\bbZ$ is given by
  $$\int_X^{\check{H}} [\check{A}]
 = \int_X \check{A} \mod \bbZ.$$
 It is just the restriction to  $\check{\cH}^{N+1}(X)$ of the integration map on the group of 
 Cheeger-Simons characters  $\check{H}^{N+1}(X)$. 
 \item Using the fundamental exact sequence of Lemma \ref{lem:exact}, we see that
 $$\check{\cH}^1(X) \cong \bbR/\bbZ \times {\cH}_\bbZ^1(X).$$
\end{enumerate}

It is well known (and easy to show) that $H^1(X, \bbZ) \cong [X, \bbR/\bbZ]$ and that 
$\check{H}^{1}(X) \cong C^\infty(X, \bbR/\bbZ)$. 
There is an analogous interpretation of
 the group of degree 1 harmonic Cheeger-Simons characters,
 given in terms of harmonic maps, defined as follows.

\begin{definition}
A smooth map $f: X \to \bbR/\bbZ$ is said to be {\em harmonic} if $\log(f): \widetilde X \to \bbR$
is a harmonic function, where $\widetilde X$ denotes the universal cover of $X$, and where 
we observe that $\gamma^*\log(f) = \log(f) + c_\gamma, \, \forall \, \gamma\in \pi_1(X)$, 
where $c_\gamma\in 2\pi\bbZ$. Let 
${\sf Harm}(X, \bbR/\bbZ)$ denote the space of all harmonic functions from $X$ to $\bbR/\bbZ$.
\end{definition}

The interpretation is then given as follows.

\begin{lemma}
Let $X$ be a compact, connected oriented Riemannian manifold. Then
$$\check{\cH}^1(X) \cong {\sf Harm}(X, \bbR/\bbZ).$$ 
\end{lemma}

\begin{proof} We show that via the isomorphism $C^\infty(X, \bbR/\bbZ) \cong \check{H}^{1}(X)$, the subgroup 
${\sf Harm}(X, \bbR/\bbZ)$ of $C^\infty(X, \bbR/\bbZ)$ corresponds to the subgroup $\check{\cH}^1(X)$
of $\check{H}^{1}(X)$. If $f \in C^\infty(X, \bbR/\bbZ)$ then consider the closed 1-form on $\widetilde X$, 
$\theta = d(\log(f))$. Now $\gamma^*\theta = d\gamma^*(\log(f)) =  d(\log(f)) = \theta$,
therefore $\theta$ descends to a 1-form $\bar\theta$ on $X$, which coincides with the fieldstrength of $f$, viewed as a differential character.

If $f \in {\sf Harm}(X, \bbR/\bbZ)$, 
$ \log(f)$ is a 
harmonic function on $\widetilde X$, and
$d^*(\theta) = d^* d(\log(f))= 0$. Therefore $\theta$ is a harmonic 
1-form on $\widetilde X$ and hence $\bar\theta$, the fieldstrength of $f$, is a harmonic 
1-form on $X$.

Conversely if $f$ has fieldstrength $\bar\theta \in \cH^1_{\ZZ}(X)$, then 
$d(\log(f)) = \pi^*\bar\theta$, where 
$\pi:\widetilde X \to X$ is the projection map. Since $\bar\theta$ is harmonic, we have
$d^*d(\log(f)) = d^* \pi^* \bar\theta = 0$. So $f \in {\sf Harm}(X, \bbR/\bbZ)$.
\end{proof}

\subsection{Poincar\'e-Pontrjagin duality}\label{subsect:PD} For a compact, connected, oriented manifold $X$,
Poincar\'e-Pontrjagin duality asserts that there is a perfect pairing defined by,
\begin{equation}\label{eqn:PPD}
\begin{aligned}
\langle.,.\rangle : &\check{\cH}^j(X) \times \check{\cH}^{N+1-j}(X) \longrightarrow \bbR/\bbZ,\\
\langle\chi_1,\chi_2\rangle &= \int_X^{\check{H}} \chi_1\star\chi_2
\end{aligned}
\end{equation}
where $\star$ is the star product on Cheeger-Simons characters. We caution the reader that 
the product of harmonic Cheeger-Simons characters is a Cheeger-Simons character,
but which may {\em not} be harmonic. This is because the field strength 
$F_{ \chi_1\star\chi_2} =  F_{\chi_1}\wedge F_{\chi_2}$ is {\em not} necessarily a harmonic form
even though $F_{\chi_1}$ and $F_{\chi_2}$ are harmonic forms. What is meant by a perfect
pairing is that every homomorphism $\check{\cH}^j(X) \longrightarrow \bbR/\bbZ$ is given 
by pairing with an element of $ \check{\cH}^{N+1-j}(X)$.

In order to justify Poincar\'e-Pontrjagin duality, we consider   two exact sequences.
The first is the fieldstrength sequence in degree $j$
\begin{equation}\label{eqn:1}
0 \to H^{j-1}(X, \bbR/\bbZ) \stackrel{\iota}{\to} \check{\cH}^j(X) \stackrel{F}{\to} \cH^j_\bbZ(X) \to 0,
\end{equation}
together with the characteristic class sequence in degree $N+1-j$,
\begin{equation}\label{eqn:2}
0 \to \cH^{N-j}(X)/\cH^{N-j}_\bbZ(X) \stackrel{I}{\to} \check{\cH}^{N+1-j}(X) \stackrel{c}{\to} H^{N+1-j}(X, \bbZ) \to 0.
\end{equation}
Recall that there is a standard perfect pairing 
$$H^{j-1}(X;\bbR/\bbZ)\times H^{N+1-j}(X;\bbZ)\to \bbR/\bbZ$$
given by multiplication and integration, and also the perfect pairing,
$$
\begin{aligned}
\cH^j_\bbZ(X) \times \left(
\cH^{N-j}(X)/\cH^{N-j}_\bbZ(X) \right)& \longrightarrow \bbR/\bbZ\\
(F, A_D) & \longrightarrow  \int_X F \wedge A_D~ \mod \bbZ .
\end{aligned}
$$
These pairings induce isomorphisms which we denote by 
\[\alpha:  H^{j-1}(X,\RR/\ZZ) \to \Hom(H^{N + 1 -j}(X,\ZZ), \RR/\ZZ)\] and 
\[\beta: \cH^j_\bbZ(X) \to \Hom(\cH^{N-j}(X)/\cH^{N-j}_\bbZ(X) , \RR/\ZZ)\] respectively. The pairing given by \eqref{eqn:PPD} induces a mapping $\varphi: \check{\cH}^j(X) \to \Hom(\check{\cH}^{N+1- j}(X), \RR/\ZZ)$. We deduce that $\varphi$ is an isomorphism as follows. The functor $\Hom( \cdot , \RR/\ZZ)$ is exact, so when applied to  \eqref{eqn:2} we obtain an exact sequence which can be combined with the exact sequence \eqref{eqn:1} into the following diagram,
\begin{eqnarray*}
\xymatrix{
0 \ar[r]&  H^{j-1}(X,\RR/\ZZ)  \ar[d]_{\alpha} \ar[r]^{\iota} & \check{\cH}^j(X) \ar[d]_{\varphi} \ar[r]^{F} & \cH^j_\bbZ(X) \ar[d]_{\beta} \ar[r]& 0 \\
 0 \ar[r]& D(H^{N + 1 -j}(X,\ZZ)) \ar[r]_{c^*} & D(\check{\cH}^{N+1- j}(X)) \ar[r]_-{I^*} & D(\cH^{N-j}(X)/\cH^{N-j}_\bbZ(X)) \ar[r] & 0,
    }
\end{eqnarray*}
where we have abbreviated $\Hom( \cdot, \RR/\ZZ)$ to $D( \cdot)$. An elementary fact from homological algebra, known as the `short five lemma', states that for any such diagram with exact rows which is commuting and where $\alpha$ and $\beta$ are isomorphisms, $\varphi$ must also be an isomorphism. Since here $\alpha$ and $\beta$ are known to be isomorphisms, it only remains to show that the diagram commutes to apply this lemma.

To see that the right square commutes, consider $\chi \in \check{\cH}^j(X)$ and $h \in \cH^{N-j}(X)$. Then it is sufficient to show that $\int_X^{\check{H}} \chi \star I(h) = \int_X F_{\chi} \wedge h \mod \ZZ$.
However the product on $\check{H}^{\bullet}(X)$ satisfies $\chi \star I(h) = I(F_\chi \wedge h)$ so
\[\int_X^{\check{H}} \chi \star I(h) = \int_X^{\check{H}} I(F_\chi \wedge h) =  \int_X F_{\chi} \wedge h \mod \ZZ.\]

Now consider $a \in H^{j-1}(X,\RR/\ZZ)$ and $\chi \in \check{\cH}^{N + 1 - j}(X)$. Then since $\iota(a)\star \chi \in \check{\cH}^{N+1}(X)$ there is a form $\tau \in \Omega^N(X)$ such that $\iota(a)\star \chi = I(\tau)$. However $\iota(a)\star \chi = \iota(a \cup c(\chi))$, so it must have vanishing fieldstrength, $d\tau = 0$,
and $\tau \mod \ZZ$ is a representative cocycle for  $a\cup c(\chi)$. Hence \[\int_X^{\check{H}} \iota(a)\star \chi = \int_X \tau \mod \ZZ = \tau(X) \mod \ZZ = (a\cup c(\chi))[X],\] and it follows that the left square in the diagram commutes.

Since the hypotheses of the `short five lemma' hold, we deduce that $\varphi$ is an isomorphism, giving 
the Poincar\'e-Pontrjagin duality \eqref{eqn:PPD}.

\subsection{Harmonic Cheeger-Simons characters on compact Lie groups}
Here we will identify harmonic Cheeger-Simons characters on a 
compact connected Lie group $G$. A Cheeger-Simons character 
$\chi$ on $G$ is said to be {\em left invariant} if $L_\gamma^*(\chi) = \chi$ for 
all $\gamma\in G$, where $L_\gamma$ denotes left translation by $\gamma \in G$.
$\chi$ is said to be {\em right invariant} if $R_\gamma^*(\chi) = \chi$ for 
all $\gamma\in G$, where $R_\gamma$ denotes right translation by $\gamma \in G$.
$\chi$ is said to be {\em bi-invariant} if it is both left invariant and right invariant.
Clearly the fieldstrength of a bi-invariant Cheeger-Simons character is a bi-invariant
differential form. Let $\check{H}^j(G)^{inv}$ denote the subgroup of all bi-invariant Cheeger-Simons
characters on $G$.

\begin{lemma}\label{lemma:harmchi}
Let $G$ be a compact connected Lie group endowed with a bi-invariant Riemannian metric. 
Then the harmonic Cheeger-Simons characters on $G$ are precisely
the bi-invariant Cheeger-Simons characters on it.
\end{lemma}

\begin{proof}
Since $G$ is connected, it acts trivially on $H^j(G,\RR/\ZZ)$, and we can choose an equivariant set theoretic splitting of the fieldstrength exact sequence. This splitting then induces an isomorphism of $G$-sets $\phi: \check{H}^j(G) \simeq H^{j-1}(G,\RR/\ZZ)\times \Omega_{\ZZ}^j(G)$, which is compatible with the fieldstrength map, $F \circ \phi = p_2$, where $p_2$ is projection onto the second factor. From this we deduce that a character is bi-invariant if and only if it has bi-invariant curvature. By a theorem of Hodge, cf. Theorem 7.8 in \cite{Helgason}, the harmonic forms on $G$ coincide with the bi-invariant differential forms on it, and so the bi-invariant characters are precisely the harmonic characters.

\end{proof}

The star product $\chi_1 \star \chi_2$ of harmonic Cheeger-Simons characters $\chi_j, \, j=1, 2,$ is 
not in general a harmonic Cheeger-Simons character, since the fieldstrength satisfies 
  $F_{\chi_1 \star \chi_2} =  F_{\chi_1}  \wedge  F_{\chi_2}$, and because it is well known that
  the wedge product of harmonic forms is not in general a harmonic form. However, in the case
  of compact connected Lie groups, we can use the Lemma \ref{lemma:harmchi} 
  above and the fact
  that the wedge product of bi-invariant forms is bi-invariant,  to deduce the following.
   
\begin{corollary}
Let $G$ be a compact connected Lie group endowed with a bi-invariant Riemannian metric. 
Then the star product $\chi_1 \star \chi_2$ of harmonic Cheeger-Simons characters $\chi_j, \, j=1, 2,$ is 
a harmonic Cheeger-Simons character.
\end{corollary}

\section{The harmonic Picard variety and the smooth geometric Hecke correspondence}\label{sect:picard}

In this section, we introduce the harmonic Picard variety, which is fundamental for 
several constructions. We also establish the geometric
Hecke correspondence for the abelian group $U(1)$ in the smooth context. This 
correspondence in the holomorphic context was established by Deligne, cf. the nice lecture 
notes \cite{Fr}.

\subsection{Geometric Hecke operators}
 Suppose that $X$ is a {compact} oriented surface with a Riemannian
metric $g$ on $X$ whose volume form is $V$ and  whose total volume 
is equal to $2\pi$. Then define the {\em harmonic Picard variety},
$$
{\check{\rm Pic}}(X, g) = \left\{ (\cL, \nabla)| F_\nabla = k V \quad \text{for some} \,\,\, k\in \Z\right\}/\sim ,
$$
where $\cL$ is a complex line bundle over $X$, $\nabla$ a connection on $\cL$
whose curvature 2-form is equal to $F_\nabla$, and $\sim$ denotes isomorphism of 
line bundles with connection. 
It is well known, cf \cite{Br, HS, FMS} and references therein, that the 2nd Cheeger-Simons
cohomology group $\check{H}^2(X)$ is isomorphic to,
$\left\{ (\cL, \nabla)\right\}/\sim ,$
where $\cL$ is a complex line bundle over $X$, $\nabla$ a connection on $\cL$, 
and $\sim$ denotes isomorphism of  line bundles with connection. 
Under this isomorphism $\check{\cH}^2(X) \subseteq \check{H}^2(X)$ corresponds to (isomorphism classes of) line bundles with connection with harmonic curvature. Since $X$ is compact and connected,  $H^2(X, \RR) \simeq \RR$, which with the Hodge decomposition implies, 
$${\check{\rm Pic}}(X, g)  \cong \check{\cH}^2(X).$$
Therefore ${\check{\rm Pic}}(X, g) $ is a finite dimensional Lie group and 
has components labelled by the integers, with ${\check{\rm Pic}}(X, g)^{(0)}$
being canonically isomorphic to the Jacobian variety ${\check{\rm Jac}(X)}$ of flat line bundles 
modulo gauge equivalence. ${\check{\rm Jac}(X)}$ is equivalently described as the space of all 
flat connections modulo gauge equivalence, 
$$
{\check{\rm Jac}}(X) = \left\{ (\cL, \nabla)| F_\nabla = 0 \right\}/\sim ,
$$
since the holonomy of a flat connection on a line bundle determines a character of the fundamental
group, and vice-versa. Since the de Rham operator $d$ is a flat connection, and since any two connections differ
by a closed 1-form, and gauge transformations can be identified with closed 1-forms with integral periods,
then we see that ${\check{\rm Jac}}(X) = H^1(X, \bbR)/H^1(X, \bbZ)$, so in particular $\pi_1({\check{\rm Jac}}(X))
= H^1(X, \bbZ)$.
Moreover, by Abelian Yang-Mills theory, one also sees that all of the other components  
${\check{\rm Pic}}(X, g)^{(n)}, \, n\in \Z\setminus \{0\}$ of ${\check{\rm Pic}}(X, g) $ are (non-canonically) isomorphic 
to ${\check{\rm Jac}(X)}$.

Consider the {\em geometric Hecke operators},
$$
\begin{array}{rcl}
{\mathbb H}_p:  {\check{\rm Pic}}(X, g)^{(n)}  & \longrightarrow & \ {\check{\rm Pic}}(X, g)^{(n+1)}\\[7pt]
(\cL , \nabla) &   \longrightarrow & ( \cL \otimes \cD_p, \nabla \otimes 1 + 1\otimes \nabla_p)
\end{array}
$$
which induce the  map,
$$
\begin{array}{rcl}
{\mathbb H} : X \times  {\check{\rm Pic}}(X, g)^{(n)} & \longrightarrow & {\check{\rm Pic}}(X, g)^{(n+1)}\\[+7pt]
			(p, (\cL, \nabla))				& \longrightarrow &{\mathbb H}_p(\cL, \nabla). 
\end{array}
$$
The main theorem of the section is,
\begin{theorem}\label{thm:geomHecke}
Given a flat line bundle $L$ over $X$, there exists a unique  flat, line bundle
$\cK_L$ over ${\check{\rm Pic}}(X, g)$, satisfying
the Hecke eigenvalue property: for $(\cL, \nabla) \in  {\check{\rm Pic}}(X, g)^{(n)}$ and $p\in X$,
$$
\cK_L\big|_{{\mathbb H}_p(\cL, \nabla)} = L_p \otimes \cK_L\big|_{(\cL, \nabla)}.
$$ 
Globally, the Hecke eigenvalue property can be rephrased as,
\begin{equation}\label{eqn:hecke}
{\mathbb H}^*\cK_L = L \boxtimes \cK_L.
\end{equation}
The Hecke correspondence is the bijection,
$$
\begin{array}{rcl}
{\check{\rm Jac}(X)}       & \mapsto & {\check{\rm Jac}}^{\rm Hecke}( {\check{\rm Pic}}(X, g)) \\[+7pt]
   L   & \mapsto & \cK_L.
\end{array}
$$
Here $ {\check{\rm Jac}}^{\rm Hecke}( {\check{\rm Pic}}(X, g))$ is the subspace 
of $ {\check{\rm Jac}}( {\check{\rm Pic}}(X, g)) $
consisting of all those flat line bundles on ${\check{\rm Pic}}(X, g)$ that satisfy the Hecke eigenvalue property
\eqref{eqn:hecke}.
\end{theorem}

\begin{proof}

It is easy to construct $\cK_L^{(0)} \in  {\check{\rm Jac}}( {\check{\rm Jac}(X)})$, as follows.
A flat line bundle $L$ over $X$ is equivalent to a character $\rho_L$ of the fundamental
group $\pi_1(X)$. Also, a character $\rho_L$ of $\pi_1(X) $ factors through a unique
character $\rho_L'$ of the Abelianization 
$\pi_1(X)_{ab} = H_1(X, \Z)$ of the fundamental group $\pi_1(X)$. Since the Riemann 
surface is oriented, Poincar\'e duality gives a natural isomorphism $ H_1(X, \Z) \cong H^1(X, \Z)
= \pi_1({\check{\rm Jac}(X)})$.
Then the definition of $\cK_L^{(0)}$ is the flat line bundle  over ${\check{\rm Jac}(X)} = {\check{\rm Pic}}(X, g)^{(0)}$ 
determined by the character $\rho_L'$ of the fundamental group $\pi_1({\check{\rm Jac}(X)})$.

We begin by defining $\cK_L$ over  ${\check{\rm Pic}}(X, g)^{(n)}$ for all $n > 0$. Let $p_j : X^n \to X$ 
denote the projection to the $j$-the factor for $j=1,\ldots ,n$. Recall that the external tensor product 
 $L^{\boxtimes n}$ is defined as the tensor product $\bigotimes_{j=1}^n p_j^*L$ of flat line 
 bundles $ p_j^*L, \, j=1,\ldots ,n$ over $X^n$, and so is also 
 a flat line bundle over $X^n$. Also, there is an action of the symmetric group $S_n$ on $L^{\boxtimes n}$
 given as follows. 
$$ 
\begin{array}{rcl}
L^{\boxtimes n} \times S_n & \longrightarrow & L^{\boxtimes n}\\[7pt]
((v_1\boxtimes \ldots \boxtimes v_n), \theta)  & \longrightarrow & (v_{\theta(1)}\boxtimes \ldots\boxtimes  v_{\theta(n)}) 
\end{array}
$$
The invariant vectors ${\sf Sym}^{(n)}(L) :=(L^{\boxtimes n})^{S_n}$ is a flat {\em orbifold} line bundle 
over the orbifold ${\sf Sym}^{(n)}(M)$,
defined by the character ${\sf Sym}^{(n)}(\rho_L)$ of the fundamental group of ${\sf Sym}^{(n)}(X)$, 
or equivalently by a character ${\sf Sym}^{(n)}(\rho_L')$ of its first homology group. 
Consider the canonical map
$h : {\sf Sym}^{(n)}(X) \times X \to {\sf Sym}^{(n+1)}(X)$ given by
$([x_1, \ldots x_n], x_{n+1})$ $ \mapsto [x_1, \ldots, x_{n+1}]$. Then clearly
\begin{equation}\label{Hecke-analogue}
h^*({\sf Sym}^{(n+1)}(L))= L \boxtimes {\sf Sym}^{(n)}(L).
\end{equation}
Therefore $ {\sf Sym}^{(n)}(L)$ satisfies the analogue of the Hecke eigenvalue property.
More generally,
\begin{equation}\label{Hecke-analogue2}
h_{m,n}^*({\sf Sym}^{(m+n)}(L))=  {\sf Sym}^{(m)}(L) \boxtimes {\sf Sym}^{(n)}(L).
\end{equation}
where $h_{m,n}$ is defined in  \eqref{eqn:symm}.

The smooth Abel-Jacobi map $A: X \to {\check{\rm Pic}}(X, g)^{(1)}$ defined by $p \mapsto [(\cD_p, \nabla_p)]$,
extends to $A_n: {\sf Sym}^{(n)}(X) \to  {\check{\rm Pic}}(X, g)^{(n)}$ given by $[p_1,\ldots, p_n]
\mapsto  [(\bigotimes_{j=1}^n \cD_{p_j}, \nabla_{p_1}\otimes 1 + \ldots 1\otimes \nabla_{p_n})]$ (notice that the last map is 
well defined since $ {\check{\rm Pic}}(X, g)$ is an Abelian group). The following diagram is commutative, showing
the compatibility of the various maps
\begin{equation}
\begin{CD}
{\sf Sym}^{(m)}(X) \times {\sf Sym}^{(n)}(X)  @> h_{m,n}>> {\sf Sym}^{(m+n)}(X) \\         
  @V{A_m \times A_n}VV     @V{A_{m+n}}VV  \\
  {\check{\rm Pic}}(X, g)^{(m)} \times   {\check{\rm Pic}}(X, g)^{(n)} @>\otimes>>   {\check{\rm Pic}}(X, g)^{(m+n)}
\end{CD}
\end{equation}
The Abel-Jacobi theorem asserts that $A_n$ induces an isomorphism,
$$H_1({\sf Sym}^{(n)}(X), \bbZ) \cong H_1({\check{\rm Pic}}(X, g)^{(n)}, \bbZ),$$
for all $n>0$, which is closely related to Proposition \ref{prop:fund-gp}, since by that proposition,
$\pi_1({\sf Sym}^{(n)}(X))$ is an abelian  group, and so by the Hurewicz theorem,
it is naturally isomorphic to $H_1({\sf Sym}^{(n)}(X), \bbZ)$.
Via this isomorphism, the character ${\sf Sym}^{(n)}(\rho_L')$  also defines a flat line bundle
$\cK_L^{(n)}$ over $ {\check{\rm Pic}}(X, g)^{(n)}$. 
Since the flat line bundle $ {\sf Sym}^{(n)}(L)$ over $ {\sf Sym}^{(n)}(X)$ 
has the analogue Hecke eigenvalue property cf equation \eqref{Hecke-analogue}, it follows that
for all  $n>0$, the Hecke eigenvalue property also holds for the flat line bundle  $\cK_L^{(n)}$  over 
$ {\check{\rm Pic}}(X, g)^{(n)}$ as claimed.

Note that, if we can define the flat line bundle $\cK_L^{(n)}$ over ${\check{\rm Pic}}(X, g)^{(n)}$ then we 
can uniquely extend $\cK_L$ to a flat line bundle  $\cK_L^{(n-1)}$
over ${\check{\rm Pic}}(X, g)^{(n-1)}$ via the Hecke eigenvalue property as follows.
By the global Hecke eigenvalue property,
${\mathbb H}^*\cK_L^{(n)}= L \boxtimes \cK_L^{(n-1)}.$
Then 
$$(L^* \boxtimes 1) \otimes {\mathbb H}^*\cK_L^{(n)} \cong (L^* \boxtimes 1) \otimes (L \boxtimes \cK_L^{(n-1)})
\cong 1 \boxtimes \cK_L^{(n-1)}$$ 
determines the flat line bundle $\cK_L^{(n-1)}$ over ${\check{\rm Pic}}(X, g)^{(n-1)}$. 
By induction, it follows that we can uniquely extend $\cK_L^{(j)}$ over $ {\check{\rm Pic}}(X, g)^{(j)}$ for all $j \le n$ and having
the Hecke eigenvalue property.

This completes the construction of the flat line bundle $\cK_L$ over ${\check{\rm Pic}}(X, g)$ having the 
Hecke eigenvalue property. 

Suppose that $\cK_L$ is the trivial bundle. Then by the Hecke eigenvalue property, it follows that $L$ is 
trivial, and therefore the Hecke correspondence is injective. On the other hand, suppose that 
$\cK \in {\check{\rm Jac}}^{\rm Hecke}( {\check{\rm Pic}}(X, g))$. Then define the flat line bundle 
$L: = \cK\big|_X$ over $X$, where $X$ is identified with its image under the (injective) Abel-Jacobi map
$A: X \to {\check{\rm Pic}}(X, g)^{(1)} \subset  {\check{\rm Pic}}(X, g)$. 
By injectivity of the Hecke correspondence, $\cK$ is equivalent to $\cK_L$.

\end{proof}

\subsection{Primitive property of flat line bundles}

The harmonic Picard variety ${\check{\rm Pic}}(X, g) $ is an Abelian Lie group under tensor 
product, 
$$
\begin{array}{rcl}
\mu : {\check{\rm Pic}}(X, g) \times {\check{\rm Pic}}(X, g) & \longrightarrow & {\check{\rm Pic}}(X, g);\\[+7pt]
\left( (\cL, \nabla) , (\cL', \nabla')\right) & \longrightarrow & [(\cL \otimes \cL', \nabla \otimes 1 + 1 \otimes \nabla')]
\end{array}
$$
The main result in this section is that any flat line bundle 
$L$ on ${\check{\rm Pic}}(X, g)$ is a {\em primitive line bundle} in the sense that it satisfies,
$$
\mu^*L \cong L \boxtimes L.
$$
Equivalently, for all elements $\cL, \cL' \in  {\check{\rm Pic}}(X, g)$, we must have:
$$
L_{\cL\otimes\cL'} \cong L_\cL \otimes L_{\cL'}.
$$
Recall that this is equivalent to the requirement that the total space of the flat principal circle bundle associated to $L$, 
is a $U(1)$-central extension of ${\check{\rm Pic}}(X, g)$.

This is but a special case of a more general result that we will now prove. 

\begin{lemma}
Let $G$ be a Lie group with fundamental group $\pi_1(G)$. Let 
$\chi : \pi_1(G) \to U(1)$ be a character of the fundamental group, and $P_\chi = \widetilde G \times_\chi U(1)$ 
be the associated flat principal $U(1)$-bundle, where $\widetilde G$ denotes the universal covering space 
of $G$, which is always also a Lie group. Then $P_\chi$ is a Lie group, which is a $U(1)$-central extension of 
$G$.
\end{lemma}

\begin{proof}
To see this, recall that $P_\chi $ consists of equivalence classes of pairs 
$(g, z) \in  \widetilde G \times U(1)$, where $(g, z) \sim (gn, \chi(n)^{-1}z)$. Consider the product 
$(g, z)(g', z') = (gg', zz')$. We need to show that this is well defined. But $(gn, \chi(n)^{-1} z)(g'n', \chi(n')^{-1} z')
= (gng'n', \chi(nn')^{-1} z z') = (gg'nn', \chi(nn')^{-1} z z')$, where we use the fact that $\pi_1(G)$ is a 
subgroup of the center of $G$. This shows that the product is well defined, and that $\{1\} \times U(1)$ is a central 
subgroup of $P_\chi$, and therefore that $P_\chi$ is a $U(1)$-central extension of 
$G$ as claimed.
\end{proof}

\begin{corollary}
For any flat line bundle $L$ over $X$, the associated flat line bundle 
$\cK_L$ is a primitive line bundle over the Abelian Lie group  ${\check{\rm Pic}}(X, g)$, i.e. it satisfies
$$
\mu^*\cK_L \cong \cK_L \boxtimes \cK_L.
$$
\end{corollary}



\begin{thebibliography}{99}

\bibitem{Br}
J-L. Brylinski,
Loop spaces, characteristic classes and geometric quantization. 
Progress in Mathematics, {\bf 107}, BirkhŠuser Boston, Inc., Boston, MA, 1993.

\bibitem{CS}
J. Cheeger and J. Simons,
Differential characters and geometric invariants. Geometry and topology, 50--80, 
Lecture Notes in Math., {\bf 1167}, Springer, Berlin, 1985.

\bibitem{dR}
G. de Rham,  
Differentiable manifolds. Forms, currents, harmonic forms. 
Grundlehren der Mathematischen Wissenschaften, {\bf 266}. Springer-Verlag, Berlin, 1984.

\bibitem{FMS}
 D. S. Freed, G. W. Moore, G. Segal,
Heisenberg Groups and Noncommutative Fluxes.
Annals Phys. {\bf 322} (2007) 236-285
[hep-th/0605200]; {\em ibid}, The uncertainty of fluxes. 
Comm. Math. Phys. {\bf 271} (2007), no. 1, 247--274. [hep-th/0605198]

\bibitem{Fr}
E. Frenkel, 
Lectures on the Langlands Program and Conformal Field Theory, 
in ``	Frontiers in Number Theory, Physics, and Geometry II'', 
editors, P. Cartier, P. Moussa, B. Julia and P. Vanhove,
pages 387--533, Springer Berlin Heidelberg, 2007.
[hep-th/0512172]. 

\bibitem{GH} 
P. Griffiths and J. Harris, 
Principles of algebraic geometry. 
Pure and Applied Mathematics. Wiley-Interscience [John Wiley \& Sons], 
New York, 1978. 

\bibitem{AG}  
A. Grothendieck, 
Revetements etales et groupe fondamental, Seminaire de 
Geometrie Algebrique du Bois-Marie (SGA 1), Lecture Notes in Math. {\bf 224}, 
Springer, Berlin, 1971. 

\bibitem{AH}
A. Hatcher,  
Algebraic topology. 
Cambridge University Press, Cambridge, 2002.

\bibitem{Helgason}
S. Helgason, 
Differential Geometry, Lie Groups, and Symmetric Spaces. 
Providence, R.I.: American Mathematical Society, 2001.

 \bibitem{HS} 
M. J. Hopkins, I. M. Singer,
Quadratic functions in geometry, topology, and M-theory. 
J. Differential Geom. {\bf 70} (2005), no. 3, 329--452.

 \bibitem{KW} 
A. Kapustin and E. Witten,
Electric-Magnetic Duality And The Geometric Langlands Program.
Commun. in Number Theory and Physics {\bf 1} (2007) 1--236.
[hep-th/0604151v3]


\bibitem{Mattuck}  
A. Mattuck, 
Symmetric products and Jacobians.
Amer. J. Math. {\bf 83} (1961) 189--206.

\end{thebibliography}
\end{document}